\documentclass[12pt]{article}
\usepackage[a4paper, total={7in, 10in}]{geometry}
\usepackage[parfill]{parskip}
\usepackage{physics, tensor, float, subcaption}
\usepackage{graphicx}
\graphicspath{ {Plots/} }
\usepackage{jhep-mod}
\usepackage{bm}
\usepackage{soul}
\usepackage{amssymb,amsmath,amsthm}
\usepackage{mathrsfs}
\usepackage[utf8]{inputenc}
\usepackage{enumerate}
\usepackage{bigints}
\usepackage{xcolor}
\usepackage{appendix}
\usepackage{graphicx}
\usepackage{float}
\usepackage{tikz}
\usepackage{setspace}
\usepackage{cancel}
\definecolor{purple}{rgb}{1,0,1}
\definecolor{lime}{HTML}{A6CE39} 

\definecolor{lime}{HTML}{A6CE39}
\newcommand{\orcidicon}{%
	\begin{tikzpicture}
	\draw[lime, fill=lime] (0,0) 
		circle [radius=0.16] 
		node[white] {{\fontfamily{qag}\selectfont \tiny ID}};
	\draw[white, fill=white] (-0.0625,0.095) 
		circle [radius=0.007];
	\end{tikzpicture}
	\hspace{-5mm}
}
\newcommand\orcidJoshua{{\href{https://orcid.org/0000-0003-1200-7261}{\orcidicon}}}
\newcommand\orcidThomas{{\href{https://orcid.org/0000-0002-0314-4136}{\orcidicon}}}
\newcommand\orcidAlex{{\href{https://orcid.org/0000-0002-1763-3563}{\orcidicon}}}
\newcommand\orcidMatt{{\href{https://orcid.org/0000-0003-1088-6485}{\orcidicon}}}

\begin{document}


\title{
Painlev\'e--Gullstrand form\\ of the Lense--Thirring spacetime
}


\author{
\Large
Joshua Baines\!\orcidJoshua\!, Thomas Berry\!\orcidThomas\!, Alex Simpson\!\orcidAlex\!,
\\{\sf  and} Matt Visser\!\orcidMatt}
\affiliation{School of Mathematics and Statistics, Victoria University of Wellington, 
\\
\null\qquad PO Box 600, Wellington 6140, New Zealand.}
\emailAdd{joshua.baines@sms.vuw.ac.nz}
\emailAdd{thomas.berry@sms.vuw.ac.nz}
\emailAdd{alex.simpson@sms.vuw.ac.nz}
\emailAdd{matt.visser@sms.vuw.ac.nz}

\abstract{
\vspace{1em}

The standard Lense--Thirring metric is a century-old slow-rotation large-distance\break \emph{approximation} to the gravitational field outside a rotating massive body, depending only on the total mass and angular momentum of the source. Although it is not an \emph{exact} solution to the vacuum Einstein equations, asymptotically the Lense--Thirring metric approaches the Kerr metric at large distances. 
Herein we shall discuss a specific \emph{variant} of the standard Lense--Thirring metric, carefully chosen for simplicity, clarity, and various forms of  ``improved'' behaviour, (to be more carefully defined in the body of the article). 
In particular we shall construct an explicit unit-lapse Painlev\'e--Gullstrand variant of the Lense--Thirring spacetime,  that has flat spatial slices, a very simple and physically intuitive tetrad, and extremely simple curvature tensors.   We shall verify that this variant of the Lense--Thirring spacetime is Petrov type I, (so it is \emph{not} algebraically special), but nevertheless possesses some very 
 straightforward timelike geodesics, (the ``rain'' geodesics). We shall also discuss on-axis and equatorial geodesics, ISCOs and circular photon orbits. Finally, we wrap up by discussing some astrophysically relevant estimates, and analyze what happens if we extrapolate down to small values of $r$.

\bigskip
\noindent
{\sc Date:} 25 June 2020; Thursday 29 October 2020; \LaTeX-ed \today


\bigskip
\noindent{\sc PhySH:} 
Gravitation
}

\maketitle
\def\tr{{\mathrm{tr}}}
\def\diag{{\mathrm{diag}}}
\def\cof{{\mathrm{cof}}}
\def\pdet{{\mathrm{pdet}}}
\parindent0pt
\parskip7pt

\clearpage
\section{Introduction}
\def\O{{\mathcal{O}}}

Only two years after the discovery of the original Schwarzschild solution in 1916~\cite{schwarzschild-1916}, in 1918 Lense and Thirring found an approximate solution to the vacuum Einstein equations at large
distances from a stationary isolated body of mass $m$ and angular momentum $J$~\cite{Lense-Thirring}.
 In suitable coordinates, at asymptotically large distances, one takes~\cite{Lense-Thirring, Pfister, Adler-Bazin-Schiffer, MTW, D'Inverno, Hartle, Carroll, kerr-intro, kerr-book}:
\begin{eqnarray}
\label{E:LT1}
d s^2 &=& - \left[1-{2m\over r} + \O\left({1\over r^2}\right)\right] \; d t^2 
- \left[{4 J \sin^2\theta\over r} +  \O\left({1\over r^2}\right)\right] \; d\phi\;d t
\nonumber
\\
&&
+ \left[1+ {2m\over r} + \O\left({1\over r^2}\right)\right] \; \left[d r^2
+ r^2 (d\theta^2+\sin^2\theta\; d\phi^2)\right].
\end{eqnarray}
Here the sign conventions are compatible  with MTW~\cite{MTW} (33.6), and Hartle~\cite{Hartle} (14.22).\break
It took another 45 years before Roy Kerr found the corresponding  exact solution in 1963~\cite{Kerr,Kerr-Texas}. 
\enlargethispage{40pt}
Nevertheless the Lense--Thirring metric continues to be of interest for two main reasons: 
(1) Lense--Thirring is \emph{much} easier to work with than the full Kerr solution; and 
(2) For a real rotating planet or star, generically possessing non-trivial mass multipole moments, the vacuum solution outside the surface is not \emph{exactly} Kerr; it is only asymptotically Kerr~\cite{kerr-intro}. (There is no Birkhoff theorem for rotating bodies in 3+1 dimensions~\cite{Birkhoff, Jebsen, Deser, Ravndal,Skakala}.) 
Consequently, the only region where one should trust the Kerr solution as applied to a real rotating star or planet is in the asymptotic regime, where in any case it reduces to the Lense--Thirring metric.

Below we shall, by suitably adjusting the sub-dominant $\O(r^{-2})$ terms,  recast a variant of the standard Lense--Thirring metric of equation (\ref{E:LT1}) into Painlev\'e--Gullstrand form --- in this form of the metric (up to coordinate transformations) one has
\begin{equation}
ds^2 = - dt^2 + \delta_{ij} (dx^i - v^i dt) (dx^j - v^j dt).
\end{equation}
That is, the constant-$t$ spatial 3-slices of the metric are all flat, and the lapse function is unity ($g^{tt}=-1$). 
See the early references~\cite{painleve1,painleve2,gullstand}, and more recently~\cite{doran,river,natario,martel,faraoni}.
(Note that the vector $v^i$, representing the ``flow'' of space, is \emph{minus} the shift vector in the ADM formalism.)
One of the virtues of putting the metric into Painlev\'e--Gullstrand form is that it is particularly easy to work with and to interpret --- in particular, the analogue spacetimes built from excitations in moving fluids are typically (conformally) of Painlev\'e--Gullstrand form~\cite{unexpected, visser:1997, visser:1998, volovik:1999, stone:2001, visser:2001, fischer:2002, novello:2002, probing, equatorial-vortex, LRR, visser:2010}, and so give a very concrete visualization of such spacetimes.

\section{Variants on the theme of the Lense--Thirring metric}
Let us now take the original Lense--Thirring metric (\ref{E:LT1}) and seek to modify and simplify it in various ways, while retaining the good features of the asymptotic large-distance behaviour.

\begin{itemize}
\item 
First, we note that at $J=0$, for a non-rotating source we do have the Birkhoff theorem so it makes sense to consider the modified metric
\begin{eqnarray}
\label{E:LT2}
d s^2 &=& - \left(1-{2m\over r}\right) \; d t^2 
- \left[{4 J \sin^2\theta\over r} +  \O\left({1\over r^2}\right)\right] \; d\phi\;d t
\nonumber
\\
&&
+{d r^2\over 1-2m/r} 
+ r^2 (d\theta^2+\sin^2\theta\; d\phi^2).
\end{eqnarray}
This modified metric asymptotically approaches standard Lense--Thirring  (\ref{E:LT1}) at large distances, but has the very strong advantage that for $J=0$ it is an \emph{exact} solution of the vacuum Einstein equations.

\item
Second, ``complete the square". Consider the modified metric
\begin{eqnarray}
\label{E:LT3}
d s^2 &=& - \left(1-{2m\over r}\right) \; d t^2 
+{d r^2\over 1-2m/r} 
\nonumber
\\
&&
+ r^2 \left(d\theta^2+\sin^2\theta\; \left(d\phi - \left[{2J\over r^3} + \O\left(1\over r^4\right)\right] dt\right) ^2\right).
\end{eqnarray}
This modified metric again asymptotically approaches standard Lense--Thirring  (\ref{E:LT1}) at large distances, but now has the \emph{two} very strong advantages that (i) for $J=0$ it is an exact solution of the vacuum Einstein equations \emph{and} (ii) that the azimuthal dependence is now in partial Painlev\'e--Gullstrand form: $g_{\phi\phi}(d\phi - v^\phi\, dt)^2= g_{\phi\phi}(d\phi-\omega dt)^2$.  See the early references~\cite{painleve1,painleve2,gullstand}, and more recently references~\cite{doran,river,natario,martel,faraoni}.

\item
Third, put the $r$--$t$ plane into standard Painlev\'e--Gullstrand form~\cite{painleve1,painleve2,gullstand,doran,river,natario,martel,faraoni}. (We note that $v^r = -  \sqrt{2m/r} $ for a Schwarzschild black hole.) We then have the modified metric
\begin{eqnarray}
\label{E:LT4}
d s^2 &=& - d t^2 +\left(d r+\sqrt{2m/r} \; dt\right)^2
\nonumber
\\
&&
+ r^2 \left(d\theta^2+\sin^2\theta\; \left(d\phi - \left[{2J\over r^3} + \O\left(1\over r^4\right)\right] dt\right) ^2\right).
\end{eqnarray}
This modified metric again asymptotically approaches standard Lense--Thirring  (\ref{E:LT1}) at large distances, but has the \emph{three} very strong advantages that (i) for $J=0$ it is an exact solution of the vacuum Einstein equations, (ii)
that the azimuthal dependence is now in partial Painlev\'e--Gullstrand form, with $g_{\phi\phi}(d\phi - v^\phi\, dt)^2 = g_{\phi\phi}(d\phi-\omega dt)^2$, 
\emph{and} (iii) that all the spatial dependence is in exact Painlev\'e--Gullstrand type form, in the sense that the constant-$t$ spatial 3-slices are now flat. 

\clearpage
\null
\vspace{-75pt}
\item
Fourth, drop the $\O(1/r^4)$ terms in the $\phi$ dependence. That is, consider the specific and fully explicit metric:
\begin{eqnarray}
\label{E:LT5}
d s^2 &=& - d t^2 +\left(d r+\sqrt{2m/r} \; dt\right)^2
+ r^2 \left(d\theta^2+\sin^2\theta\; \left(d\phi - {2J\over r^3} dt\right) ^2\right).\qquad
\end{eqnarray}
By construction for $J=0$ this is the Painlev\'e--Gullstrand version of the Schwarzschild metric~\cite{painleve1,painleve2,gullstand,doran,river,natario}.
By construction at large distances this asymptotically approaches the ``standard" form of Lense--Thirring as given in equation (\ref{E:LT1}), and so it also asymptotically approaches Kerr. 
By construction even for $J\neq0$ this metric is in Painlev\'e--Gullstrand form. 
(In particular, with flat spatial 3-slices, and as we shall soon see, unit lapse, and easily constructed timelike geodesics.) These observations make this specific form (\ref{E:LT5}) of the Lense--Thirring spacetime particularly interesting and worth investigation.
\end{itemize}
We emphasize that the five spacetimes represented by these five metrics (\ref{E:LT1})--(\ref{E:LT2})--(\ref{E:LT3})--(\ref{E:LT4})--(\ref{E:LT5}) are physically different from each other.  They may have the same asymptotic limit at large distances, but differ in many crucial technical details. In particular, as we shall soon see, the tetrads, curvature components, and the analysis of geodesics is much easier for the fully explicit  Painlev\'e--Gullstrand form of the metric (\ref{E:LT5}) than it is for any of the (\ref{E:LT1})--(\ref{E:LT2})-- (\ref{E:LT3})--(\ref{E:LT4}) variants. 

\vspace{-10pt}
\section{Metric components, tetrad, and curvature}
\vspace{-10pt}

\enlargethispage{40pt}
We shall now analyze the metric, a particularly natural choice of tetrad (vierbein), the curvature invariants, and the orthonormal tetrad components of the curvature tensors for our Painlev\'e--Gullstrand variant of the Lense--Thirring spacetime.

\subsection{Metric components}

From (\ref{E:LT5}) it is easy to read off the metric components
\begin{equation}
g_{ab} = \left[ \begin{array}{c|ccc}
-1+{2m\over r} + {4 J^2\sin^2\theta\over r^4} &  \sqrt{2m\over r} & 0 & -{2J\sin^2\theta\over r}\\
\hline
\sqrt{2m\over r} & 1 & 0 & 0\\
0& 0 & r^2 & 0\\
-{2J\sin^2\theta\over r} & 0 & 0 & r^2\sin^2\theta\\
\end{array}
\right]_{ab}.
\end{equation}
\enlargethispage{40pt}
Thence one can easily verify that for the inverse metric
\begin{equation}
\label{E:contra_metric}
g^{ab} = \left[ \begin{array}{c|ccc}
-1&  \sqrt{2m\over r} & 0 & -{2J\over r^3}\\
\hline
\sqrt{2m\over r} & 1-{2m\over r} & 0 & \sqrt{2m\over r} \;{2J\over r^3}\\
0& 0 & {1\over r^2} & 0\\
-{2J\over r^3} &  \sqrt{2m\over r}\; {2J\over r^3} & 0 & {1\over r^2\sin^2\theta} - {4J^2\over r^6}\\
\end{array}
\right]^{ab}.
\end{equation}
Note particularly that $g^{tt}=-1$, so that the lapse function is unity; this fact will be particularly useful when we come to analyzing the geodesics.
\bigskip

\clearpage
\subsection{Tetrad}

Let us denote tetrad labels by an overhat:  That is set $\hat a, \hat b \in\{\hat t, \hat r,\hat\theta,\hat\phi\}$.  Furthermore set $\eta_{\hat a \hat b} = \mathrm{diag}(-1,1,1,1)$. 
To find a suitable covariant  tetrad (co-tetrad) $e^{\hat a}{}_a$, we wish to find a particular solution of $g_{ab} = \eta_{\hat a\hat b} \; e^{\hat a}{}_a\; e^{\hat b}{}_b$. 
Then from the line-element (\ref{E:LT5}) an obvious and straightforward choice for the co-tetrad is
\begin{eqnarray}
e^{\hat t}{}_a &=& (1;0,0,0); \qquad 
\; \; e^{\hat r}{}_a = \left(\sqrt{2m\over r};1,0,0\right); 
\nonumber\\[2pt]
e^{\hat \theta}{}_a&=& r(0;0,1,0); \qquad 
e^{\hat \phi}{}_a = r\sin\theta\left(-{2J\over r^3};0,0,1\right).
\end{eqnarray}
This choice of co-tetrad is of course not unique. (The underlying metric  is unaffected by any arbitrary local Lorentz transformation $L^{\hat a}{}_{\hat b}$ on the ortho-normal tetrad/co-tetrad indices.) However this co-tetrad is particularly well-adapted to the coordinate system used in (\ref{E:LT5}). 
Once the co-tetrad has been chosen, the contravariant tetrad (usually just called the tetrad) is then uniquely defined by $e_{\hat a}{}^a = \eta_{\hat a\hat b} \; e^{\hat b}{}_b \;g^{ba}$.

The tetrad therefore will satisfy 
\begin{eqnarray}
\eta^{\hat a\hat b} \; e_{\hat a}{}^a\; e_{\hat b}{}^b &=& 
\eta^{\hat a\hat b} \;  (\eta_{{\hat a \hat c}} \; e^{\hat c}{}_c \;g^{ca}) (\eta_{{\hat b\hat d}} \; e^{\hat d}{}_d \;g^{da}) = 
\eta_{{\hat c\hat d}}\;  (e^{\hat c}{}_c \;g^{ca}) (e^{\hat d}{}_d \;g^{da}) 
\nonumber\\
&=& g_{cd} \; g^{ca} \; g^{db} = g^{ab}.
\end{eqnarray}
A brief computation, or comparison with the inverse metric (\ref{E:contra_metric}), leads to
\begin{eqnarray}
e_{\hat t}{}^a &=& \left(1;-\sqrt{2m\over r},0,{2J\over r^3}\right); \qquad 
\; \; e_{\hat r}{}^a = \left(0;1,0,0\right); 
\nonumber\\[2pt]
e_{\hat \theta}{}^a&=& {1\over r}\;(0;0,1,0); \qquad 
e_{\hat \phi}{}^a = {1\over r\sin\theta}\left(0;0,0,1\right).
\end{eqnarray}
Note that the last 3 of these tetrad vectors are exactly those that would be expected for flat Euclidean 3-space, and that for this choice of tetrad all of the nontrivial physics is tied up in the timelike vector $e_{\hat t}{}^a$. 
For our purposes the tetrad and co-tetrad are most usefully employed in converting tensor coordinate components into an orthonormal basis.

\clearpage
\subsection{Curvature invariants}

While the specific Lense--Thirring spacetime variant we are interested in, that of equation (\ref{E:LT5}), is not (exactly) Ricci-flat, it is easy to calculate the Ricci scalar and Ricci invariant and verify that asymptotically they are suitably small. We have
\begin{equation}
R = \displaystyle{18 J^2 \sin^2\theta\over r^6};
\end{equation}
and
\begin{equation}
R_{ab} \,R^{ab} = 3 R^2.
\end{equation}
Note that all the right things happen as $J\to 0$.
Note that all the right things happen as $r\to \infty$.
Ultimately, it is the observation that these quantities fall-off very rapidly with distance that justifies the assertion that this is an ``approximate'' solution to the vacuum Einstein equations.

A more subtle calculation is to evaluate the Weyl invariant:
\begin{eqnarray}
C_{abcd} \,C^{abcd} 
&= & 
 {48 m^2\over r^6}  -{144 J^2(2\cos^2\theta+1) \over r^8}   +{864mJ^2\sin^2\theta\over r^9}
 +{1728 J^4 \sin^4\theta\over r^{12}}\qquad
 \nonumber \\
 &= & 
 {48 m^2\over r^6}  -{144 J^2(3-2\sin^2\theta) \over r^8}   +{48m\over r^3} \; R 
 +{16\over 3} \; R^2\qquad
\nonumber \\
 &=&
 {48 m^2\over r^6}  -{432 J^2\over r^8} + {16\over r^2} \left(1+{3m\over r}\right) \; R 
 +{16\over 3} \; R^2.
\end{eqnarray}
Note that this is exactly what you would expect for Schwarzschild, $48m^2/r^6$, plus a rapid fall-off  angular-momentum-dependent term, $\O(J^2/ r^8)$. 
Similarly for the Kretschmann scalar we have
\begin{equation}
R_{abcd} R^{abcd} = C_{abcd} \,C^{abcd}  + {1728 J^4 \sin^4\theta\over r^{12}} 
=  C_{abcd} \,C^{abcd} + {17\over3} \; R^2.
\end{equation}

\subsection{Curvature tensors}

Calculating the Ricci and Einstein tensors is (in the tetrad basis) straightforward.
\begin{itemize}
\item 
Taking $R_{\hat a\hat b} = e_{\hat a}{}^a \; e_{\hat b}{}^b \; R_{ab}$, in terms of the Ricci scalar $R$  we have:
\begin{equation}
R_{\hat a\hat b}= R\left[ \begin{array}{c|ccc}
-1&0& 0 &0\\ \hline  0& 1 & 0 & 0\\
0 & 0 & 0 & 0 \\ 0& 0 & 0 &  -1
\end{array}\right].
\end{equation}
Notice the perhaps somewhat unexpected pattern of zeros and minus signs.

\item 
Taking $G_{\hat a\hat b} = e_{\hat a}{}^a \; e_{\hat b}{}^b \; G_{ab}$, in terms of the Ricci scalar $R$  we have:
\begin{equation}
G_{\hat a \hat b} = {R\over2} \left[ \begin{array}{c|ccc}
-1&0& 0 &0\\ \hline  0& 1 & 0 & 0\\
0 & 0 & -1 & 0 \\ 0& 0 & 0 &  -3
\end{array}\right].
\end{equation}
Notice the perhaps somewhat unexpected pattern of zeros and minus signs.
\enlargethispage{40pt}
\item
Algebraically, this implies that the Ricci and Einstein tensors are type I in the Hawking--Ellis (Segre--Plebanski)
classification~\cite{prado1,prado2}.
\end{itemize}
In contrast, calculating the Weyl and Riemann tensors is somewhat tedious.
\begin{itemize}
\item 
Take $C_{\hat a\hat b\hat c\hat d } = e_{\hat a}{}^a \; e_{\hat b}{}^b \; e_{\hat c}{}^c \; e_{\hat d}{}^d \; C_{abcd}$.
\begin{itemize}
\item 
The terms quadratic in $J$ are:
\begin{eqnarray}
C_{\hat t\hat r \hat t \hat r } 
&=& -2 C_{\hat t\hat \theta \hat t \hat \theta }  = -2 C_{\hat t\hat \phi \hat t \hat \phi } 
= 2 C_{\hat r\hat \theta \hat r \hat \theta }  = 2 C_{\hat r\hat \phi \hat r \hat \phi } 
= - C_{\hat \theta \phi \hat \theta \hat\phi}
\nonumber\\
&=& -{2m\over r^3} -{12 J^2\sin^2\theta\over r^6}= -{2m\over r^3} -{2\over 3} R.
\end{eqnarray}
\item
There are also several terms linear in $J$:
\begin{eqnarray}
{1\over2} C_{\hat t \hat r \hat \theta \hat\phi} &=& C_{\hat t \hat \theta \hat r \hat\phi} =
- C_{\hat t \hat \phi \hat r \hat\theta}
= {3J\cos\theta\over r^4};
\nonumber\\
C_{\hat t \hat r \hat r \hat\phi} &=& -C_{\hat t \hat \theta \hat \theta \hat\phi} 
=-{3J\sin\theta\over r^4};
\nonumber\\
C_{\hat t \hat r \hat t \hat\phi}&=& - C_{\hat r \hat \theta \hat \theta \hat\phi}
= {3J\sin\theta \sqrt{2m/r}\over r^4}.
\end{eqnarray}
\end{itemize}
\item
Take $R_{\hat a\hat b\hat c\hat d } = e_{\hat a}{}^a \; e_{\hat b}{}^b \; e_{\hat c}{}^c \; e_{\hat d}{}^d \; R_{abcd}$.
\begin{itemize}
\item 
 The terms quadratic in $J$ are:
\begin{eqnarray}
R_{\hat t\hat r \hat t \hat r } 
&=& -{2m\over r^3} -{27 J^2\sin^2\theta\over r^6} =
-{2m\over r^3} -{3\over2} R.
\nonumber\\
R_{\hat t\hat \phi \hat t \hat \phi } &=& - R_{\hat r\hat \phi \hat r \hat \phi } 
= {m\over r^3} +{9 J^2\sin^2\theta\over r^6} 
=  {m\over r^3}+ {1\over2} R.
\end{eqnarray}
\item
There are also several terms linear in $J$:
\begin{eqnarray}
R_{\hat t \hat r \hat \theta \hat\phi} &=& 2R_{\hat t \hat \theta \hat r \hat\phi} =
- 2R_{\hat t \hat \phi \hat r \hat\theta}
= {6J\cos\theta\over r^4};
\nonumber\\
R_{\hat t \hat r \hat r \hat\phi} &=& -R_{\hat t \hat \theta \hat \theta \hat\phi} 
=-{3J\sin\theta\over r^4};
\nonumber\\
R_{\hat t \hat r \hat t \hat\phi}&=& - R_{\hat r \hat \theta \hat \theta \hat\phi}
= {3J\sin\theta \sqrt{2m/r}\over r^4}.
\end{eqnarray}
\item
There are now also several terms independent of $J$:
\begin{eqnarray}
R_{\hat t \hat \theta\hat t \hat\theta} &=& -R_{\hat r \hat \theta \hat r \hat\theta} =
{1\over2}  R_{\hat \theta \hat \phi \hat \theta \hat\phi}
= {m\over r^3};
\end{eqnarray}
\end{itemize}
\end{itemize}
Overall, the tetrad components of the Weyl and Riemann tensors are quite tractable; the coordinate components are considerably more complicated.

\section{Petrov type I}

It is straightforward (if somewhat tedious) to check that the Painlev\'e--Gullstrand version of the Lense--Thirring metric is Petrov type I. (That is, the Lense--Thirirng geometry is \emph{not algebraically special}.)
To do this, one proceeds by first calculating the mixed Weyl tensor 
$C^{\hat a\hat b}{}_{\hat c\hat d}$.
Now since this object is antisymmetric in the individual pairs $[\hat a\hat b]$ and $[\hat c\hat d]$ this can effectively be thought of as a real $6\times6$ matrix according to the scheme 
$A \longleftrightarrow [\hat a\hat b]$ and $B \longleftrightarrow [\hat c\hat d]$ as follows: 
\[
1 \leftrightarrow [\hat1\hat2]; \qquad 2 \leftrightarrow [\hat1\hat3]; \qquad 
3 \leftrightarrow [\hat1\hat4]; \qquad 4 \leftrightarrow [\hat3\hat4]; \qquad 
5 \leftrightarrow [\hat4\hat2]; \qquad 6 \leftrightarrow [\hat2\hat3].
\]
Note that this $6\times 6$ matrix $C^A{}_B$ is not symmetric,  nor should it be symmetric.

\enlargethispage{10pt}
It is particularly useful to first define the two real quantities
\begin{equation}
\Xi_1=- {m\over r^3} - {6J^2\sin^2\theta\over r^6}; \qquad 
\Xi_2 = {3J\cos\theta\over r^4}.
\end{equation}
Further defining $s=\sin(\theta)$ the $6\times 6$ matrix $C^A{}_B$ is:
\begin{equation}
C^A{}_B=
\left[
\begin{array}{ccc|ccc}
%
-2\,\Xi_1 & 0 &-3\sqrt{\frac{2m}{r}}{Js\over r^4} & -2\,\Xi_2 & -{3Js\over r^4} & 0\\
%
0 & \Xi_1 & 0 & -{3Js\over r^4} & \Xi_2 & 0\\
%
-3\sqrt{\frac{2m}{r}}{Js\over r^4} & 0& \Xi_1 & 0 & 0 & \Xi_2\\
\hline
2\,\Xi_2 &  {3Js\over r^4}  & 0 & -2\,\Xi_1 & 0& -3\sqrt{\frac{2m}{r}}{Js\over r^4} \\
%
 {3Js\over r^4}  & -\Xi_2 & 0 & 0 & \Xi_1 & 0\\
 %
0 & 0 & -\Xi_2 &-3\sqrt{\frac{2m}{r}}{Js\over r^4}  & 0 & \Xi_1
\end{array}
\right].
\end{equation}
Note that the $6\times 6$ matrix $C^A{}_B$ is traceless, $C^A{}_A=0$. (As it must be since $C^{ab}{}_{ab}=0$.)
This asymmetric matrix nevertheless has the partial symmetry
\begin{equation}
C^A{}_B= \left[ \begin{array}{c|c} S_R & S_I\\  \hline - S_I & S_R \end{array} \right].
\end{equation}
Here $S_R$ and $S_I$ are themselves symmetric traceless $3\times3$ matrices.

\clearpage
There are 6 distinct eigenvalues, 
appearing in complex conjugate pairs. Explicitly, 
defining the third real quantity
\begin{equation}
\Xi_3 = \left( 3Js\over r^4 \right)^2 - \left(3\sqrt{\frac{2m}{r}}{Js\over r^4} \right)^2
 = 9\left(1-{2m\over r}\right) {J^2 \sin^2\theta\over r^8},
\end{equation}
the 6 eigenvalues of $C^A{}_B$ are
\begin{equation}
\label{E:weyl_plus}
\Xi_1 + i\,\Xi_2; \qquad  
-{1\over2} (\Xi_1 + i\,\Xi_2) \pm \sqrt{ {\textstyle{9\over 4}} (\Xi_1 + i\,\Xi_2)^2- \Xi_3};
\end{equation}
and
\begin{equation}
\label{E:weyl_minus}
\Xi_1 - i\,\Xi_2; \qquad  \qquad
-{1\over2} (\Xi_1 - i\,\Xi_2) \pm \sqrt{ {\textstyle{9\over 4}} (\Xi_1 - i\,\Xi_2)^2-  \Xi_3}.
\end{equation}
Note that the sum over all six eigenvalues yields zero, as it must do since the underlying matrix $C^A{}_B$ is traceless.

The fact that there are (generically) 6 distinct eigenvalues guarantees that the Jordan canonical form of $C^A{}_B$ is trivial, and therefore that the spacetime is of Petrov type I.
(That is, the Painlev\'e--Gullstrand from of Lense--Thirring is \emph{not algebraically special}.)
Relevant discussion can be found on pages 49 and 50 of the ``Exact solutions'' book by Stephani \emph{et al}~\cite{exact}.
Note that those authors prefer to rearrange the $6\times 6$ real matrix $C^A{}_B$ into  a pair of $3\times3$ complex matrices:
\begin{equation}
S_\pm = S_R \pm i S_I =
\left[\begin{array}{ccc}
-2(\Xi_1\pm i \Xi_2) &  \mp{3iJs\over r^4} & -3\sqrt{\frac{2m}{r}}{Js\over r^4}\\
 \mp{3iJs\over r^4}  & \Xi_1\pm i \Xi_2 & 0\\
-3\sqrt{\frac{2m}{r}}{Js\over r^4} & 0 &\Xi_1\pm i \Xi_2
\end{array}\right].
\end{equation}
This really makes no difference to the physics of the discussion, but does simplify the matrix algebra.  
It is easy to check that the 3 eigenvalues of $S_+$ are given by (\ref{E:weyl_plus}), and that the 3 eigenvalues of $S_-$ are given by (\ref{E:weyl_minus}). 
In view of the fact that $\tr(S_\pm)=0$, each set of 3 eigenvalues must independently sum to zero, as they explicitly do.

Let us now consider some special cases:
\begin{itemize}
\item 
On the rotation axis we have $\theta\to 0$. So $\Xi_1\to-{m\over r^3}$, while $\Xi_2\to{3J\over r^4}$ and $\Xi_3\to 0$. Then $S_\pm$ are both complex diagonal and the 6 Weyl eigenvalues collapse to 
\begin{equation}
\{\lambda\} \to \left\{  -{m\over r^3}\pm {3iJ\over r^4},\; 
 -{m\over r^3}\pm {3iJ\over r^4};
 -2 \left( -{m\over r^3}\pm {3iJ\over r^4}\right)  \right\}.
\end{equation}
So on-axis the 6 Weyl eigenvalues \emph{are} degenerate.

\clearpage
\item
On the equator we have  $\theta\to \pi/2$. So $\Xi_1\to-{m\over r^3}-{6J^2\over r^6}$, while $\Xi_2\to0$ and finally $\Xi_3\to 9(1-{2m\over r}) J^2/r^8$. Then $S_+=S_-$ and the eigenvalues collapse to the twice-repeated degenerate values
\begin{equation}
\{\lambda\} \to \left\{ 
\Xi_1,  
-{1\over2} \Xi_1 \pm \sqrt{ {\textstyle{9\over 4}} (\Xi_1)^2-\Xi_3}
\right\}.
\end{equation}
\item
Finally note that when  $J\to0$ we have $\Xi_1\to-{m\over r^3}$, while $\Xi_2\to0$ and  $\Xi_3\to 0$. Then $S_\pm$ are both real and diagonal and the 6 Weyl eigenvalues collapse to 
\[
\{\lambda\} 
\to
 \left\{  -{m\over r^3},\;-{m\over r^3},\; -{m\over r^3}, \;-{m\over r^3}; \; \;
  {2m\over r^3},\;  {2m\over r^3} \right\}.
\]
This is exactly the repeated eigenvalue structure you would expect for the Schwarzschild spacetime. 
\end{itemize}
That is: While the Weyl eigenvalues are degenerate on-axis, on the equator, and in the non-rotating $J\to0$ limit, the generic situation is that there are six distinct eigenvalues;  our Painlev\'e--Gullstrand variant of Lense--Thirring is Petrov type I. 
While, generally speaking, Petrov type I is normally associated with a \emph{lack} of 
special properties, we shall soon see that the Painlev\'e--Gullstrand variant of Lense--Thirring still has many very nice features when it comes to the analysis of geodesics. 

\section{``Rain'' geodesics}

For our Painlev\'e--Gullstrand variant of Lense--Thirring spacetime at least some of the timelike geodesics, the ``rain'' geodesics corresponding to a test object being dropped from spatial infinity with zero initial velocity and zero angular momentum, are particularly easy to analyze. (These are someimes called ZAMOs --- zero angular momentum observers.) Consider the vector field
\begin{equation}
V^a = - g^{ab} \nabla_b t = - g^{ta} = \left(1;-\sqrt{2m\over r}, 0, {2J\over r^3}\right).
\end{equation}
This implies
\begin{equation}
V_a = -  \nabla_a t =  \left(-1;0, 0,0\right).
\end{equation}
Thence $g_{ab} V^a V^b = V^a V_a = -1$, so $V^a$ is a  future-pointing timelike vector field with unit norm, a 4-velocity. But then this vector field has zero 4-acceleration:
\begin{equation}
A_a = V^b \nabla_b V_a = - V^b \nabla_b \nabla_a t = - V^b \nabla_a \nabla_b t 
=  V^b \nabla_a V_b = {1\over2} \nabla_a (V^b V_b) = 0.
\end{equation}
Thus the integral curves of $V^a$ are timelike \emph{geodesics}. 
For this construction to work it is \emph{essential} that the metric be unit-lapse --- so while this works nicely for our (\ref{E:LT5}) Painlev\'e--Gullstrand variant of Lense--Thirring, it will \emph{fail} for any and all of the (\ref{E:LT1})--(\ref{E:LT2})-- (\ref{E:LT3})--(\ref{E:LT4}) variants of Lense--Thirring spacetime.

Specifically, the  integral curves represented by
\begin{equation}
{dx^a\over d\tau} = \left( {dt\over d\tau};{dr\over d\tau},{d\theta\over d\tau},{d\phi\over d\tau}\right) =
\left( 1 ;  -\sqrt{2m/r}, 0, 2J/r^3\right) 
\end{equation}
are timelike geodesics.
Integrating two of these equations is trivial
\begin{equation}
t(\tau) = \tau; \qquad \theta(\tau)=\theta_\infty;
\end{equation}
so that the time coordinate $t$ can be identified with the proper time of these particular geodesics, and $\theta_\infty$ is the original (and permanent) value of the $\theta$ coordinate for these particular geodesics. 

Furthermore, algebraically one has 
\begin{equation}
{1\over2} \left(dr\over dt\right)^2 = {m\over r};
\end{equation}
so these particular geodesics mimic Newtonian infall from spatial infinity with initial velocity zero.

Finally note that
\begin{equation}
{d\phi\over dr} = {d\phi/d\tau\over dr/d\tau} = -{2J/r^3\over \sqrt{2m/r}} = -{2J\over \sqrt{2m}} \; r^{-5/2},
\end{equation}
which is easily integrated to yield
\begin{equation}
\phi(r) = \phi_\infty +{4J\over3\sqrt{2m}} \; r ^{-3/2}.
\end{equation}
Here $\phi_\infty$ is the initial value of the $\phi$ coordinate (at $r=\infty$) for these particular geodesics.
Note the particularly clean and simple way in which rotation of the source causes these ``rain'' geodesics to be deflected. 
These pleasant features are specific to our (\ref{E:LT5}) Painlev\'e--Gullstrand variant of Lense--Thirring, and \emph{fail} for the (\ref{E:LT1})--(\ref{E:LT2})-- (\ref{E:LT3})--(\ref{E:LT4}) variants of Lense--Thirring spacetime.

\section{On-axis geodesics}

Working on-axis we have either $\theta\equiv0$ or $\theta\equiv\pi$, and so $\dot\theta=0$. Working on-axis we can, without loss of generality,  also choose $\dot\phi=0$. Then we need only consider the $t$--$r$ plane, and the specific variant of the Lense--Thirring metric that we are interested in effectively reduces to
\begin{eqnarray}
\label{E:LT-axis}
d s^2 &\to& - d t^2 +\left(d r+\sqrt{2m/r} \; dt\right)^2.
\end{eqnarray}
That is, we effectively have
\begin{equation}
g_{ab} \to \left[ \begin{array}{c|c}
-1+{2m\over r} &  \sqrt{2m\over r} \\
\hline
\sqrt{2m\over r} & 1 \\
\end{array}
\right];
\qquad
g^{ab} \to \left[ \begin{array}{c|c}
-1&  \sqrt{2m\over r} \\
\hline
\sqrt{2m\over r} & 1-{2m\over r} \\
\end{array}
\right].
\end{equation}
This observation is enough to guarantee that on-axis the geodesics of our specific (\ref{E:LT5}) Painlev\'e--Gullstrand variant of the  Lense--Thirring spacetime  are \emph{identical} to those for the Painlev\'e--Gullstrand version of the Schwarzschild spacetime. 
(For a related discussion, see for instance the discussion by Martel and Poisson in reference~\cite{martel}.) 
For the on-axis null curves $x^a(t)=(t,r(t))$ we have $g_{ab} \; (dx^a/dt)\; (dx^b/dt) =0 $ implying
\begin{equation}
-1 + \left( {dr\over dt} + \sqrt{2m/r} \right)^2 =0.
\end{equation}
That is, for on-axis null curves (as expected for a black hole) we have
\begin{equation}
\label{E:null-on-axis}
{dr\over dt} =  - \sqrt{2m\over r} \pm 1.
\end{equation}

For on-axis timelike geodesics we parameterize by proper time $x^a(\tau)=(t(\tau),r(\tau))$. 
Then we have $g_{ab} \; (dx^a/d\tau)\; (dx^b/d\tau) =-1 $, implying
\begin{equation}
\left(dt\over d\tau\right)^2 \left(-1 + \left( {dr\over dt} + \sqrt{2m/r} \right)^2 \right) =-1.
\end{equation}
From the time translation Killing vector $K^a=(1;0,0,0)^a\to(1,0)^a$ we construct the conserved quantity:
\begin{equation}
K_a \; (dx^a/d\tau) = k.
\end{equation}
Thence
\begin{equation}
\left(dt\over d\tau\right) \left( \left(-1+{2m\over r}\right) + \sqrt{2m\over r} \; {dr\over dt} \right) = k.
\end{equation}
Eliminating $dt/d\tau$ we see
\begin{equation}
k^2 \left(-1 + \left( {dr\over dt} + \sqrt{2m/r} \right)^2 \right) = 
-  \left( \left(-1+{2m\over r}\right) + \sqrt{2m\over r} \; {dr\over dt} \right)^2.
\end{equation}
This is a quadratic for $dr/dt$, with explicit general solution
\begin{equation}
{dr\over dt} = -\sqrt{2m\over r} \;\; {k^2-1+2m/r\over k^2+2m/r} 
\pm k \;{\sqrt{k^2-1+2m/r}\over k^2+2m/r}.
\end{equation}
The limit $k\to\infty$ reproduces the result for on-axis null geodesics given in (\ref{E:null-on-axis}).

\enlargethispage{40pt}
As $r\to\infty$ one has
\begin{equation}
\lim_{r\to\infty} \left(dr\over dt\right) = \pm \sqrt{1-{1\over k^2} },
\end{equation}
which provides a physical interpretation for the parameter $k$. Indeed
\begin{equation}
k = {1\over\sqrt{1- \left(dr\over dt\right)_\infty^2}}
\end{equation}
is the asymptotic ``gamma factor'' of the on-axis geodesic (which may be less than unity, and $\left(dr\over dt\right)_\infty$ might formally be imaginary,  if the geodesic is bound).
As $k\to 1$ the negative root corresponds to the ``rain'' geodesic falling in from spatial infinity with zero initial velocity, so that $dr/dt = - \sqrt{2m/r}$, while the positive root yields 
\begin{equation}
{dr\over dt} = \sqrt{2m\over r} \left(1 -2m/r \over 1+2m/r\right).
\end{equation}
This represents an outgoing timelike geodesic with ${dr\over dt}$ asymptoting to zero at large distances.
Overall, the on-axis geodesics of our variant Lense-Thirring spacetime are quite simple to deal with.

\section{Generic non-circular equatorial geodesics}

For equatorial geodesics we set $\theta=\pi/2$, and consequently $\dot\theta=0$. 
For generic non-circular equatorial geodesics it proves most efficient to work directly in terms of the conserved Killing quantities associated with the timelike and azimuthal Killing vectors. (For circular equatorial geodesics, discussed in the next section, the effective potential proves to be a more useful tool.)
Working on the equator we need only consider the $t$--$r$--$\phi$ hypersurface, and our specific (\ref{E:LT5}) Painlev\'e--Gullstrand variant of the Lense--Thirring metric effectively reduces to
\begin{eqnarray}
\label{E:LT-equator}
d s^2 &\to& - d t^2 +\left(d r+\sqrt{2m/r} \; dt\right)^2
+ r^2  \left(d\phi - {2J\over r^3} dt\right) ^2.
\end{eqnarray}
That is, we effectively have
\begin{equation}
g_{ab} \to \left[ \begin{array}{c|cc}
-1+{2m\over r} + {4 J^2\over r^4} &  \sqrt{2m\over r} \;\; & -{2J\over r}\\
\hline
\sqrt{2m\over r} & 1 & 0\\
-{2J\over r} & 0 & r^2\\
\end{array}
\right],
\end{equation}
and thence
\begin{equation}
g^{ab} \to \left[ \begin{array}{c|cc}
-1&  \sqrt{2m\over r} & -{2J\over r^3}\\
\hline
\sqrt{2m\over r} & 1-{2m\over r} & \sqrt{2m\over r} \;{2J\over r^3}\\
-{2J\over r^3} &  \sqrt{2m\over r}\; {2J\over r^3}&\; {1\over r^2} - {4J^2\over r^6}\\
\end{array}
\right].
\end{equation}

\subsection{Equatorial non-circular null geodesics}

For equatorial (non-circular) null geodesics let us parameterize the geodesic curve $x^a(\lambda)= (t(\lambda);x^i(\lambda))$  by some arbitrary affine parameter $\lambda$. Then the null condition $g_{ab} \; (dx^a/dt)\; (dx^b/dt) =0 $ implies 
\begin{equation}
\label{E:null}
 - 1 +\left({d r\over dt} +\sqrt{2m/r} \right)^2
+ r^2  \left({d\phi\over dt} - {2J\over r^3}\right) ^2=0.
\end{equation}
From the time translation and azimuthal Killing vectors, $K^a=(1;0,0,0)^a\to(1;0,0)^a$ and $\tilde K^a=(0;0,0,1)^a\to(0,0,1)^a$, we construct the two conserved quantities:
\begin{equation}
K_a \; \left(dx^a\over d\lambda\right)  = k; 
\qquad \hbox{and} \qquad 
\tilde K_a \; \left(dx^a\over d\lambda\right) =  \tilde k. 
\end{equation}
Explicitly these yield
\begin{equation}
{dt\over d\lambda} \left(-1+{2m\over r} +{4J^2\over r^4} 
+ \sqrt{2m\over r} \; {dr \over dt} - {2J\over r} \; {d\phi\over dt} \right) = k,
\end{equation}
and 
\begin{equation}
{dt\over d\lambda} \left( -{2J\over r} + r^2 \;{d\phi\over dt} \right) = \tilde k.
\end{equation}
Eliminating $dt/d\lambda$ between these two equations we see
\begin{equation}
\tilde k \left(-1+{2m\over r} +{4J^2\over r^4} 
+ \sqrt{2m\over r} \; {dr \over dt} - {2J\over r} \; {d\phi\over dt} \right) 
= k r^2\left({d\phi\over dt}  -{2J\over r^3}\right).
\end{equation}
This can be solved, either for $d\phi/dt$ or for $dr/dt$, and then substituted back into the null condition (\ref{E:null}) to yield a quadratic, either for $dr/dt$ or for $d\phi/dt$.
These quadratics can be solved, exactly, for $dr/dt$ or for $d\phi/dt$, but the explicit results are messy. 
Recalling that the Lense--Thirring spacetime is at least in its original incarnation  a large-distance approximation, 
it makes sense to peel off the leading terms in an expansion in terms of inverse powers of $r$.

\enlargethispage{20pt}
For $dr/dt$ one then finds 
\begin{equation}
{dr \over dt} = - \sqrt{2m\over r} \; P(r) \pm  \sqrt{Q(r)},
\end{equation}
where $P(r)$ and $Q(r)$ are rational polynomials in $r$ that asymptotically satisfy
\begin{equation}
P(r) = 1 - {\tilde k^2\over k^2 r^2} + \O(1/r^5); 
\qquad
Q(r)= 1 - \left(1+{2m\over r}\right) {\tilde k^2\over k^2 r^2}  
+  \O(1/r^5).
\end{equation}
Fully explicit formulae for $P(r)$ and $Q(r)$ can easily be found but are quite messy to write down.

Similarly for $d\phi/dt$ one finds
\begin{equation}
{d\phi\over dt} =  \left({2J\over r^3}- {\tilde k\over k r^2}  \right) \tilde P(r) \pm \sqrt{2m\over r}\;  {\tilde k\over k r^2} \; \sqrt{\tilde Q(r)}\,.
\end{equation}
Here $\tilde P(r)$ and $\tilde Q(r)$ are rational polynomials in $r$ that asymptotically satisfy
\begin{equation}
\tilde P(r) = 1 - {2 \tilde k (Jk+m\tilde k) \over  k^2 r^3} + \O(r^{-4}); 
\qquad
\tilde Q(r)= 1 - {\tilde k^2\over k^2 r^2} - {2 \tilde k (2Jk+m\tilde k) \over  k^2 r^3}
+  \O(r^{-5}).
\end{equation}
Fully explicit formulae for $\tilde P(r)$ and $\tilde Q(r)$ can easily be found but are quite messy to write down.
Overall, while equatorial null geodesics are in principle integrable, they are in practice not entirely tractable.

\subsection{Equatorial non-circular timelike geodesics}

For equatorial timelike geodesics the basic principles are quite similar. First let us parameterize the curve $x^a(\tau)$ using the proper time parameter. Then the timelike normalization condition $g_{ab} \; (dx^a/d\tau)\; (dx^b/d\tau) =-1 $ implies 
\begin{equation}
\label{E:timelike}
\left(dt\over d\tau\right)^2 \left(  - 1 +\left({d r\over dt} +\sqrt{2m/r} \right)^2
+ r^2  \left({d\phi\over dt} - {2J\over r^3}\right) ^2 \right)=-1.
\end{equation}
From the time translation and azimuthal Killing vectors, $K^a=(1;0,0,0)^a\to(1;0,0)^a$ and $\tilde K^a=(0;0,0,1)^a\to(0,0,1)^a$, we construct the two conserved quantities:
\begin{equation}
K_a \; \left(dx^a\over d\tau\right)  = k; 
\qquad \hbox{and} \qquad 
\tilde K_a \; \left(dx^a\over d\tau\right) =  \tilde k. 
\end{equation}
Explicitly these yield
\begin{equation}
{dt\over d\tau} \left(-1+{2m\over r} +{4J^2\over r^4} 
+ \sqrt{2m\over r} \; {dr \over dt} - {2J\over r} \; {d\phi\over dt} \right) = k,
\end{equation}
and 
\begin{equation}
\label{E:angular}
{dt\over d\tau} \left(- {2J\over r} + r^2 \;{d\phi\over dt} \right) = \tilde k.
\end{equation}
Eliminating $dt/d\tau$ between these two equations we see
\begin{equation}
\label{E:linear}
\tilde k \left(-1+{2m\over r} +{4J^2\over r^4} 
+ \sqrt{2m\over r} \; {dr \over dt} - {2J\over r} \; {d\phi\over dt} \right) 
= k r^2 \left( {d\phi\over dt}- {2J\over r^3}\right).
\end{equation}
Eliminating $dt/d\tau$ between (\ref{E:angular}) and (\ref{E:timelike}) we see
\begin{equation}
\label{E:timelike2}
\tilde k^2 \left(  - 1 +\left({d r\over dt} +\sqrt{2m/r} \right)^2
+ r^2  \left({d\phi\over dt} - {2J\over r^3}\right) ^2 \right)=
-\left( -{2J\over r} + r^2 \;{d\phi\over dt} \right)^2.
\end{equation}
Equation (\ref{E:linear}) can be solved, either for $d\phi/dt$ or for $dr/dt$, and then substituted back into the modified timelike normalization condition (\ref{E:timelike2}) to yield a quadratic, 
either for $dr/dt$ or for $d\phi/dt$. 
As for the null geodesics, it is useful to work perturbatively at large $r$.

Fot $dr/dt$ one then finds 
\begin{equation}
{dr \over dt} = - \sqrt{2m\over r} \;  \; P(r) \pm \sqrt{Q(r)},
\end{equation}
where $P(r)$ and $Q(r)$ are rational polynomials in $r$ that asymptotically satisfy
\begin{equation}
P(r) = 1 - k^{-2} + {2m\over k^4 r} + \O(1/r^2); 
\qquad
Q(r)= 1 - k^{-2} +  {2m(2-k^2)\over k^4 r} +  \O(1/r^2).
\end{equation}
Fully explicit formulae for $P(r)$ and $Q(r)$ can easily be found but are quite messy to write down.

Similarly for $d\phi/dt$ one finds
\begin{equation}
{d\phi\over dt} =  \left({2J\over r^3}  -{\tilde k\over k r^2}  \right) \tilde P(r) 
\pm \sqrt{2m\over r}\;  {\tilde k\over k r^2} \; \sqrt{\tilde Q(r)} 
\end{equation}
where $\tilde P(r)$ and $\tilde Q(r)$ are rational polynomials in $r$ that asymptotically satisfy
\begin{equation}
\tilde P(r) = 1 - {2 m \over  k^2 r} + \O(1/r^2); 
\qquad
\tilde Q(r)= 1 - k^{-2} + {2m(2-k^2)\over k^4 r}  +  \O(1/r^2).
\end{equation}
Fully explicit formulae for $\tilde P(r)$ and $\tilde Q(r)$ can easily be found but are quite messy to write down.
Overall, while equatorial non-circular timelike geodesics are in principle integrable, they are in practice not entirely tractable.

\section{Circular equatorial geodesics}

For circular equatorial geodesics the use of the effective potential formalism proves to be most efficient.
Recall that the line element for our variant of the Lense--Thirring spacetime is:
\begin{eqnarray}
\label{E:LT5b}
d s^2 &=& - d t^2 +\left(d r+\sqrt{\frac{2m}{r}} \; dt\right)^2
+ r^2 \left(d\theta^2+\sin^2\theta\; \left(d\phi - {2J\over r^3} dt\right) ^2\right) .
\end{eqnarray}
Now consider the tangent vector to the worldline of a massive or massless particle, parameterized by some arbitrary affine parameter, $\lambda$:
\begin{eqnarray}
    g_{ab}\frac{dx^{a}}{d\lambda}\frac{dx^{b}}{d\lambda} &=& -\left(\frac{dt}{d\lambda}\right)^{2} + \left[\left(\frac{dr}{d\lambda}\right)+\sqrt{\frac{2m}{r}}\left(\frac{dt}{d\lambda}\right)\right]^{2} \nonumber \\
    && \nonumber \\
    && + r^{2}\left\lbrace\left(\frac{d\theta}{d\lambda}\right)^{2} +\sin^{2}\theta\left[\left(\frac{d\phi}{d\lambda}\right)-\frac{2J}{r^{3}}\left(\frac{dt}{d\lambda}\right)\right]^{2}\right\rbrace \ .
\end{eqnarray}
We may, without loss of generality, separate the two physically interesting cases (timelike and null) by defining:
\begin{equation}
    \epsilon = \left\{
    \begin{array}{rl}
    -1 & \qquad\mbox{massive particle, \emph{i.e.} timelike worldline} \\
     0 & \qquad\mbox{massless particle, \emph{i.e.} null worldline} .
    \end{array}\right. 
\end{equation}
That is, ${ds^{2}\over d\lambda^2}=\epsilon$. We now consider geodesics on the equatorial plane, that is, we fix $\theta=\frac{\pi}{2}$ (hence, $\frac{d\theta}{d\lambda}=0$). These geodesics now represent (not yet circular) orbits restricted to the equatorial plane \emph{only}. The timelike/null condition now reads:
\begin{equation}\label{epsilon}
    -\left(\frac{dt}{d\lambda}\right)^{2} + \left[\left(\frac{dr}{d\lambda}\right)+\sqrt{2m\over r}\left(\frac{dt}{d\lambda}\right)\right]^{2} + r^{2}\left[\left(\frac{d\phi}{d\lambda}\right)-\frac{2J}{r^{3}}\left(\frac{dt}{d\lambda}\right)\right]^{2} = \epsilon \ .
\end{equation}

\enlargethispage{40pt}

\noindent The Killing symmetries in the $t$ and $\phi$--coordinates yield the following expressions for the conserved energy $E$ and angular momentum $L$ per unit mass:
\begin{eqnarray}\label{EL}
    E &=& \left(-1+\frac{2m}{r}+\frac{4J^{2}}{r^{4}}\right)\left(\frac{dt}{d\lambda}\right) + \sqrt{\frac{2m}{r}}\left(\frac{dr}{d\lambda}\right) - \frac{2J}{r}\left(\frac{d\phi}{d\lambda}\right) \ ; \\
    && \nonumber \\
    \label{EG} L &=& r^{2}\left(\frac{d\phi}{d\lambda}\right) - \frac{2J}{r}\left(\frac{dt}{d\lambda}\right) \ .
\end{eqnarray}
Treating equations~(\ref{epsilon}), (\ref{EL}), (\ref{EG}) as a system of three equations
 in the three unknowns $\frac{dt}{d\lambda}$, $\frac{dr}{d\lambda}$, and $\frac{d\phi}{d\lambda}$, we can rearrange and solve for $\frac{dr}{d\lambda}$ as a function of the metric parameters, $r$, $E$, $L$, and $\epsilon$ \emph{only}. This process yields:
\begin{equation}
{dr\over d\lambda} = \pm \sqrt{ 
\left(E+ {2JL\over r^3}\right)^2 - \left(1-{2m\over r}\right)\left({L^2\over r^2}-\epsilon\right)} \ .
\end{equation}
We can now solve for the effective potential and then use the features of this effective potential to solve for the radial positions of the circular photon orbits and innermost stable circular orbits of our spacetime. The potential is given by:
\begin{equation}
V(r) =  E^2-  \left(\frac{dr}{d\lambda}\right)^2 
=\left(1-{2m\over r}\right) \left({L^2\over r^2}-\epsilon\right) 
+E^2 - \left( E +{2LJ\over r^3}\right)^2
\end{equation}
Notice that in the limit where $J\rightarrow 0$, the potential is manifestly that of Schwarzschild. We now consider two separate cases, the massless case where $\epsilon=0$ and the massive case where $\epsilon=-1$. We shall start our analysis with the massless case. 

\subsection{Circular null orbits}

In the massless case where $\epsilon=0$, our potential reduces to 
\begin{equation}
\begin{split}
V_0(r) & =  \left(1-{2m\over r}\right)  {L^2\over r^2}
 - {4EJL\over r^3} - {4 L^2 J^2\over r^6}\\
& =   {L( L(r-2m)r^3 + 4J(JL+Er^3))\over r^6}.
\end{split}
\end{equation}
The photon ring of our spacetime occurs where $\frac{dV_0(r)}{dr}=0$. That is, the value of $r$ where the following condition is met:
\begin{equation}
\frac{dV_0(r)}{dr} = -{2L\over r^7} \left(L r^4 - 3(Lm+2EJ) r^3 -12 J^2 L  \right) = 0.
\end{equation}
This quartic equation has no tractable analytic solution. However, a more tractable semi-analytic solution can be obtained if we solve the two equations $V_0(r)=E^2$ and $\frac{dV_0(r)}{dr}=0$ simultaneously. That is, we solve the following polynomials simultaneously for~$r$:
\begin{equation}
L r^4 - 3(Lm+2EJ) r^3 -12 J^2 L = 0 \, ;
\end{equation}
\begin{equation}
E^2 r^6 - L^2 r^4 +2L(Lm+2EJ)r^3+ 4 J^2L^2=0.
\end{equation}
If we eliminate $E$ from these equations, we find
\begin{equation}
r^5-6mr^4+9mr^3+72J^2m-36rJ^2=0.
\end{equation}
(Notice that $L$ has also been eliminated in this process). Now we rearrange:
\begin{equation}
r^3(r-3m)^2 = 36J^2(1-2m/r) r.
\end{equation}
Thence
\begin{equation}
r = 3m \pm {6J\sqrt{1-2m/r}\over r}.
\end{equation}
This is still exact. To now estimate the value of $r$ corresponding to the location of the photon ring purely in terms of the parameters present in our spacetime, we iterate the lowest-order estimate $r = 3m + \O(J)$ to yield
\begin{equation}
r = 3m \pm {6J\sqrt{1-2/3}\over 3m} + \O(J^2).
\end{equation}
Finally
\begin{equation}
r = 3m \pm {2J\over \sqrt{3} \; m} + \O(J^2).
\end{equation}
Notice that in the limit where $J\rightarrow 0$, the photon ring reduces to its known location in Schwarzschild. Also, note that in the Kerr geometry, the photon ring for massless particles occurs at:
\begin{equation}
r_{\text{Kerr}}=2m\left[ 1+\cos\left(\frac{2}{3}\cos^{-1}\left(\pm \frac{J}{m^2}\right)\right)\right].
\end{equation}
If we conduct a Taylor series expansion around $J=0$, we find
\begin{equation}
\begin{split}
r_{\text{Kerr}}(J\rightarrow 0) & =2m\left(\frac{3}{2}\pm \frac{J}{\sqrt{3}\;m^2} + \O(J^2)\right)\\
& = 3m \pm {2J\over \sqrt{3} \; m} + \O(J^2),
\end{split}
\end{equation}
which is exactly the photon ring location in the Lense--Thirring spacetime. This shows that in the slow-rotation limit, the Kerr solution does reduce to Lense--Thirring as we expect. 

\noindent In terms of stability of these orbits, we analyse the second derivative of the potential $V_0(r)$:
\begin{equation} \label{ddV0}
\frac{d^2V_0(r)}{dr^2}=- {6L\over r^8} \left( 28 J^2 L +4(mL+2EJ) r^3-Lr^4\right).
\end{equation}
However, here we cannot simply eliminate $L$ as we did before. Instead we solve $\frac{dV_0(r)}{dr}=0$ for $L$, which gives
\begin{equation}
L=- {6 EJ r^3\over 12J^2+3mr^3-r^4}.
\end{equation}
Substituting this back into equation \eqref{ddV0} we find
\begin{equation}
\frac{d^2V_0(r)}{dr^2} = - {72 E^2 J^2 (r^4+36 J^2)\over r^2 (12J^2+3mr^3-r^4)^2},
\end{equation}
which is everywhere negative. Hence all equatorial circular null geodesics in our Painlev\'e--Gullstrand variant of the Lense--Thirring spacetime are unstable.

\subsection{ISCOs}

In the massive case (timelike orbits) where $\epsilon=-1$, our potential reduces to:
\begin{equation}
V_{-1}(r) =  \left(1-{2m\over r}\right) \left(1+{L^2\over r^2}\right) - {4EJL\over r^3} - {4 L^2 J^2\over r^6}.
\end{equation}
Taking the derivative of this
\begin{equation}
\frac{dV_{-1}(r)}{dr} = {2\over r^7} \left(m r^5 - L(L r^4 - 3(Lm+2EJ) r^3 -12 J^2 L).  \right)
\end{equation}
Similarly to the null case, $\frac{dV_{-1}(r)}{dr}=0$, which is now a quintic,  has no analytic solution. However, we can begin to form an analytic solution if we solve both $V_{-1}(r)=E^2$ and $\frac{dV_{-1}(r)}{dr}=0$ simultaneously. That is, we solve the following polynomials simultaneously for r:
\begin{equation}
m r^5 - L(L r^4 - 3(Lm+2EJ) r^3 -12 J^2 L) =0;
\end{equation}
\begin{equation} \label{E_quad}
 (E^2-1)r^6 +2m r^5 - L^2 r^4 +2L(Lm+2EJ)r^3 + 4 J^2 L^2 =0.
\end{equation}
If we extract $E$ from the first of these equations, we find
\begin{equation}
E=- {mr^5 - L^2 r^4+3mL^2 r^3+12J^2L^2\over 6JLr^3}.
\end{equation}
Substituting this back into $\eqref{E_quad}$ we find the following condition:
\begin{equation} \label{Circ_Con_T}
r^3(3L^2m-L^2r+mr^2)^2  -36J^2L^2(L^2+r^2)(r-2m)=0.
\end{equation}
This condition shows that there exist \emph{many} circular timelike orbits $r(L,J,m)$. Unlike the null case, $L$ is not eliminated, hence we cannot solve for the ISCO location yet. We next find the second derivative of the potential:
\begin{equation}
\frac{d^2V_{-1}(r)}{dr^2} = -{2\over r^8} \left( 2m r^5 -3 L^2 r^4 +12L(mL+2EJ)r^3+84J^2L^2\right).
\end{equation}
Now substituting our expression for $E$:
\begin{equation}
\frac{d^2V_{-1}(r)}{dr^2} = -{2\over r^8} (L^2(r^4+36J^2) -2m r^5).
\end{equation}
The condition for an extremal equatorial circular orbit is $\frac{d^2V_{-1}(r)}{dr^2}=0$, that is:
\begin{equation}
(L^2(r^4+36J^2) -2m r^5) =0.
\end{equation}
Using this condition and our condition for an equatorial circular orbit, equation (\ref{Circ_Con_T}), we can now eliminate $L$ and hence find
\begin{equation}
m r^6 (r-6m)^2 = 72 J^2 (r^2+mr-10m^2) + 1296J^4(2r-5m).
\end{equation}
This implicitly defines $r(m,J)$ in terms of the mass and angular momentum of the spacetime.
Thence, rearranging
\begin{equation}
(r-6m)^2 = {72 J^2 r^3(r^2+mr-10m^2) + 1296J^4(2r-5m)\over m r^6}.
\end{equation}
Thence, finally
\begin{equation}
r=6m \pm {6J\over m r^3} \sqrt{ 2r^3 (r^2+mr-10m^2) + 36J^2(2r-5m)}.
\end{equation}
This is still exact. However, to estimate the value of $r(m,J)$ corresponding to the location of the ISCO purely in terms of the parameters present in our spacetime, we iterate the zeroth-order estimate $r = 6m +\O(J)$ to yield
\begin{equation}
r = 6m +  {4\sqrt{2}\over\sqrt{3}}{J\over m} +  \O(J^2).
\end{equation}
Notice that in the limit where $J\rightarrow 0$, the ISCO reduces to its known location in Schwarzschild. Also, note that in the Kerr geometry, the ISCO for massive particles occurs at:
\begin{equation}
r_{\text{Kerr}} = m \left(3 + Z_2\pm\sqrt{(3-Z_1)((3+Z_1+2Z_2)}\right).
\end{equation}
Here
\begin{equation}
x = J/m^2; \qquad Z_1 = 1 + \sqrt[3]{(1-x^2)}\left(\sqrt[3]{(1+x)} +\sqrt[3]{(1-x)}\right); \qquad 
Z_2 = \sqrt{3x^2+Z_1^2}.
\end{equation}
If we conduct a Taylor series expansion around $J=0$ we find 
\begin{equation}
r_{\text{Kerr}}(J\rightarrow 0) = 6m +  {4\sqrt{2}\over\sqrt{3}}{J\over m} +  \O(J^2).
\end{equation}
which is exactly the ISCO location in the Lense--Thirring spacetime. This shows that in the slow-rotation limit, the Kerr solution does reduce to Lense--Thirring as we expect.

\section{Astrophysically relevant estimates}

Note that in SI units 
\begin{equation}
m = {G_N \; m_{physical}\over c^2}; \qquad J = {G_N \; J_{physical}\over c^3}.
\end{equation}
So dimensionally
\begin{equation}
[m] = [\hbox{length}]; \qquad [J] = [\hbox{length}]^2.
\end{equation}
It is also useful to introduce the quantities $a=J/m$ and $a/m=J/m^2$ so that
\begin{equation}
[a] = [J/m] =  [\hbox{length}]; \qquad [a/m] = [J/m^2] = [\hbox{dimensionless}].
\end{equation}
For uncollapsed objects (stars, planets) we may proceed by \emph{approximating} the source as a constant-density rigidly rotating sphere of radius $R_{source}$, angular velocity $\omega$, and equatorial velocity $v_{equatorial}$. In the Newtonian approximation
\begin{equation}
J_{physical} = I \,\omega =   {2\over 5}\,m_{physical} \; R_{source}^2 \; \omega = {2\over 5}\,m_{physical} \; R_{source} \; v_{equatorial}.
\end{equation}
Thence in geometrodynamic units we have the approximations
\begin{equation}
J = {2\over 5}m \; R_{source} \; {v_{equatorial}\over c};\qquad
a = {J\over m} = {2\over 5} R_{source} \; {v_{equatorial}\over c}.
\end{equation}
Furthermore, (defining $r_{Schwarzschild}=2m$ in geometrodynamic units),
\begin{equation}
{a\over m} = {J\over m^2} = {4\over 5} \; {R_{source}\over r_{Schwarzschild}} \; {v_{equatorial}\over c}.
\end{equation}
Another useful dimensionless parameter is
\begin{equation}
{J\over R_{source}^2 } = {1\over 5} \; {r_{Schwarzschild} \over R_{source} } \; {v_{equatorial}\over c}.
\end{equation}
Using this discussion it is possible to \emph{estimate} the parameters $m$,  $J$,  $a=J/m$, $a/m=J/m^2$ and $J/R_{source}^2$ for various astrophysically interesting objects such as the Earth, Jupiter, Sun, Sagittarius A$^*$, the black hole in M87, and our own Milky Way galaxy. See table~\ref{T:1} for details.

\clearpage
\begin{table}[h!]
\caption{Some astrophysical estimates.}
\label{T:1}
{\small
\hspace{-35pt}
\begin{tabular}{|c||c|c|c|c|c|c|}
\hline\hline
Source & $m$ (metres) & $J$ (metres)$^2$ & $a$ (metres) & $J/m^2$ (dimensionless)  & $J/R_{source}^2$ \\
\hline\hline
Earth 
&0.004435& 0.01755 & 3.959 & 892.5 & $4.315\times 10^{-16}$ \\
Jupiter 
& 1.409& 1615& 1415&812.9&$3.304\times 10^{-13}$\\
Sun 
& $1477$ & $2.741 \times 10^{6}$ & $1855$ & $1.256$ & $5.652 \times 10^{-12}$\\
\hline
Sagittarius A$^*$ 
& $6.5 \times 10^9 $& $1.9\times10^{19}$ &  $2.9 \times 10^9$ & $\approx 0.44$  & $\approx 0.12$ \\
Black hole in M87 & 
$3.5\times 10^{12}$ & $1.1\times 10^{25}$& $3.2\times10^{12}$&$\approx 0.90$& $\approx 0.44$ \\
\hline
Milky Way galaxy 
& $1.5\times 10^{15}$ &$2.5\times 10^{31}$& $1.7\times 10^{16}$ & $\approx 11$ & $\approx 10^{-10}$ \\
\hline\hline
\end{tabular}
}
\end{table}

\enlargethispage{30pt}
To interpret the physical significance of table~\ref{T:1}, first note that Kerr black holes in standard Einstein gravity must satisfy $a/m<1$, that is $J/m^2<1$,  in order to avoid development of naked singularities. However no such constraint applies to uncollapsed objects. Observationally, we do seem to have $J/m^2<1$ for the object Sagittarius A$^*$ and the central object in M87, (which are indeed believed to be Kerr black holes, at least approximately), while $J/m^2 > 1$ for the Earth, Jupiter, Sun, and the Milky Way galaxy. 

Secondly, observe that the Lense--Thirring metric should (in its original asymptotic form) really only be applied in the region $r>R_{source}$, and  for uncollapsed sources we certainly have $J/R_{source}^2 \ll 1$. Even for collapsed sources we still see $J/R_{source}^2 \lesssim 1$.
The fact that the dimensionless number $J/R_{source}^2 \ll 1$ for the Earth, Jupiter, Sun, (and even the Milky Way galaxy), is an indication that Lense--Thirring spacetime is a perfectly good approximation for the gravitational field generated by these sources \emph{once one gets beyond the surface of these objects}. 

These observations are potentially of interest when studying various black hole mimickers~\cite{phenomenology, pandora, complete, viability,LISA}.
(To include a spherically symmetric halo of dark matter in galactic sources, simply replace $m\to m(r)$. The existence of the gravitationally dominant dark matter halo is really the only good reason for treating spiral galaxies as approximately spherically symmetric.)

\section{Singularity, horizon, ergo-surface, and the like}

Now recall that the original motivation for considering the Lense--Thirring metric really only makes sense for $r>R_{source}$. 
In fact the Lense--Thirring metric is likely to be a good approximation to the exterior spacetime geometry only for $J/r^2 \ll 1$, that is $r \gg \sqrt{J}$.
But one can nevertheless ask, (both for pedagogical purposes and with a view to exploring potential black hole mimickers), what happens if we extrapolate our variant of the Lense--Thirring metric  down to $r\to 0$, and investigate the possible occurrence of horizons and ergo-surfaces.

\clearpage
Indeed, extrapolating our variant of the Lense--Thirring spacetime down to $r= 0$ one sees that there is a point curvature singularity at $r=0$.
Furthermore, note that $\nabla_a r$ becomes timelike for $r<2m$.
That is, $g^{ab} \;\nabla_a  r \; \nabla_b r = g^{rr} = 1 -{2m\over r}$, and this changes sign at $r=2m$.
Thence for $r<2m$ any future-pointing timelike vector must satisfy $V^a \;\nabla_a r <0$.
That is, (in contrast to the Kerr spacetime), there is a single horizon at the Schwarzschild radius $r=2m$, an outer horizon with no accompanying inner horizon.
Finally, note that one cannot ``stand still'' once $g_{tt}<0$.
That is, the time-translation Killing vector becomes spacelike once $ g_{ab} \,K^a \,K^b = g_{tt} >0$ corresponding to 
\begin{equation}
1 - {2m\over r} - {4J^2 \sin^2\theta\over r^4} <0.
\end{equation}
That is, there is an ergo-surface located at
\begin{equation}
r_E(\theta)^4 - {2m \; r_E(\theta)^3} - {4J^2 \sin^2\theta} =0.
\end{equation}
That is,
\begin{equation}
r_E(\theta) = 2m + {4J^2 \sin^2\theta\over r_E(\theta)^3}.
\end{equation}

On axis we have $r_E(\theta=0)=r_E(\theta=\pi) = 2m$, so that on axis the ergo-surface touches the horizon at $r_H=2m$.
Near the axis, (more precisely for $J\sin^2\theta/ m^2 \ll 1$),  the formula for $r_E(\theta)$ can be perturbatively solved  to yield
\begin{equation}
r_E(\theta) = 2m \left\{1 +{J^2\sin^2\theta\over 4 m^4} - {3J^4\sin^4\theta\over 16m^8} 
+ \O\left( J^6\sin^6\theta\over m^{12} \right) \right\}.
\end{equation}
At the equator we have either
\begin{equation}
r_E(\theta=\pi/2) = 2m \left\{1 +{J^2\over 4 m^4} - {3J^4\over 16m^8} 
+ \O\left( J^6\over m^{12} \right) \right\},
\end{equation}
or
\begin{equation}
r_E(\theta=\pi/2) = \sqrt{2J}\left\{ 1 + {m\over 2\sqrt{2J}} + {3 m^2 \over 16 J} 
 + \O\left(m^3\over J^{3/2} \right) \right\},
\end{equation}
depending on whether $J \ll m^2$ or $J \gg m^2$. 

Generally we have a quartic to deal with, while there is an exact solution it is so complicated as to be effectively unusable, and the best we can analytically say is to place the simple and tractable lower bounds
\begin{equation}
r_E(\theta) > \max\left\{ 2m, \sqrt{2J\sin\theta} \right\},
\end{equation}
and
\begin{equation}
r_E(\theta) >   \sqrt[4]{(2m)^4 + 4J^2 \sin^2\theta}. 
\end{equation}
For a tractable upper bound we note
\begin{equation}
r_E(\theta) = 2m + {4J^2 \sin^2\theta\over r_E(\theta)^3} < 2m + {4J^2 \sin^2\theta\over (2m)^3},
\end{equation}
whence
\begin{equation}
r_E(\theta) < 2m\left\{ 1 + {J^2 \sin^2\theta\over 4 m^4} \right\} < 2m\left\{ 1 + {J^2\over 4 m^4} \right\}.
\end{equation}

Overall, if one does extrapolate our variant of the Lense--Thirring spacetime down to $r= 0$, one finds a point singularity at $r=0$, a horizon at the Schwarzschild radius, 
and an ergo-surface at $r_E <  2m\left\{ 1 + {J^2\over 4 m^4} \right\}$. 
While such extrapolation is astrophysically inappropriate for vacuum spacetime in standard general relativity, it may prove interesting for pedagogical reasons, or for exploring additional examples of potential black-hole mimickers.

\section{Conclusions}

What have we learned form this discussion?

First, the specific variant of the Lense--Thirring spacetime given by the metric 
\begin{eqnarray}
\label{E:LT5b}
d s^2 &=& - d t^2 +\left(d r+\sqrt{2m/r} \; dt\right)^2
+ r^2 \left(d\theta^2+\sin^2\theta\; \left(d\phi - {2J\over r^3} dt\right) ^2\right)
\end{eqnarray}
is a very tractable and quite reasonable model for the spacetime region exterior to rotating stars and planets. 
Because this metric is in Painlev\'e--Gullstrand form, the physical interpretation is particularly transparent. 
Furthermore, with the slight generalization $m\to m(r)$, that is, 
\begin{eqnarray}
\label{E:LT5b}
d s^2 &=& - d t^2 +\left(d r+\sqrt{2m(r)/r} \; dt\right)^2
+ r^2 \left(d\theta^2+\sin^2\theta\; \left(d\phi - {2J\over r^3} dt\right) ^2\right)\quad
\end{eqnarray}
one can accommodate spherically symmetric dark matter halos, so one has a plausible approximation to the gravitational fields of spiral galaxies. Best of all, this specific Painlev\'e--Gullstrand variant of the  Lense--Thirring spacetime  is rather easy to work with. 

\section*{Acknowledgements}

\enlargethispage{10pt}
JB was supported by a MSc scholarship funded by the Marsden Fund, 
via a grant administered by the Royal Society of New Zealand.

TB was supported by a Victoria University of Wellington MSc scholarship, 
and was also indirectly supported by the Marsden Fund, 
via a grant administered by the Royal Society of New Zealand.

AS was supported by a Victoria University of Wellington PhD Doctoral Scholarship,
and was also indirectly supported by the Marsden fund, 
via a grant administered by the Royal Society of New Zealand.

MV was directly supported by the Marsden Fund, 
via a grant administered by the Royal Society of New Zealand.



\begin{thebibliography}{99}




\bibitem{schwarzschild-1916}
K. Schwarzschild,
``\"Uber das Gravitationsfeld eines Massenpunktes nach der Einsteinschen Theorie'',
Sitzungsberichte der K\"oniglich Preussischen Akademie der Wissenschaften \textbf{7} (1916) 189.
\href{https://tinyurl.com/y99c33n5}{Free online version}.


\bibitem{Lense-Thirring}
Hans Thirring and Josef Lense,  ``\"Uber den Einfluss der Eigenrotation der Zentralk\"orperauf die Bewegung 
der Planeten und Monde nach der Einsteinschen Gravitationstheorie'', Physikalische Zeitschrift, Leipzig Jg. {\bf 19}  (1918), No. 8, p. 156--163.\\ 
English translation by Bahram Mashoon, Friedrich W. Hehl, and Dietmar S. Theiss: ``On the influence of the proper rotations of central bodies on the motions of planets and moons in Einstein's theory of gravity'', General Relativity and Gravitation  {\bf 16} (1984) 727--741.

\bibitem{Pfister}
Herbert Pfister, ``On the history of the so-called Lense--Thirring effect'',
{\color{blue} \sf http://philsci-archive.pitt.edu/archive/00002681/01/lense.pdf}

\bibitem{Adler-Bazin-Schiffer}
Ronald J. Adler, Maurice Bazin, and Menahem Schiffer, \\
 {\sl Introduction to General Relativity}, Second edition, \\
 (McGraw--Hill, New York, 1975).\\
 {}[It is important to acquire the 1975 second edition, the 1965 first edition does not contain any discussion of the Kerr spacetime.]

\bibitem{MTW}
Charles Misner, Kip Thorne, and John Archibald Wheeler,  {\sl Gravitation}, \\
(Freeman, San Francisco, 1973).

\bibitem{D'Inverno}
Ray D'Inverno, {\sl Introducing Einstein's Relativity}, (Oxford University Press, 1992).

\bibitem{Hartle}
James Hartle, {\sl Gravity: An introduction to Einstein's general relativity},\\
(Addison Wesley, San Francisco, 2003).

\bibitem{Carroll}
Sean Carroll, {\sl  An introduction to general relativity: Spacetime and Geometry},
(Addison Wesley, San Francisco, 2004).

\bibitem{kerr-intro}
M.~Visser,
``The Kerr spacetime: A brief introduction'',
[\href{https://arxiv.org/abs/0706.0622}{arXiv:0706.0622 [gr-qc]}].
Published in \cite{kerr-book}.


\clearpage
\bibitem{kerr-book}
D.~L.~Wiltshire, M.~Visser and S.~M.~Scott (editors),\\
\emph{The Kerr spacetime: Rotating black holes in general relativity},\\
(Cambridge University Press, Cambridge, 2009).




\bibitem{Kerr}
Roy Kerr, 
``Gravitational field of a spinning mass as an example of algebraically special metrics'',
 Physical Review Letters {\bf  11} 237-238 (1963).

\bibitem{Kerr-Texas}
Roy Kerr, ``Gravitational collapse and rotation'', published in: 
{\sl Quasi-stellar sources and gravitational collapse:
Including the proceedings of the First Texas Symposium on Relativistic
Astrophysics}, edited by Ivor Robinson, Alfred Schild, and E.L. Sch\"ucking
(University of Chicago Press, Chicago, 1965), pages 99--102.\\
The conference was held in Austin, Texas, on 16--18 December 1963.



\bibitem{Birkhoff}
Garret Birkhoff,  
 \emph{Relativity and Modern Physics}, 
(Harvard University Press, Cambridge, 1923).

\bibitem{Jebsen}
J\o{}rg Tofte Jebsen, ``\"Uber die allgemeinen kugelsymmetrischen 
L\"osungen der Einsteinschen Gravitationsgleichungen im Vakuum'', 
Ark. Mat. Ast. Fys. (Stockholm) {\bf 15} (1921) nr.18. 

\bibitem{Deser}
  Stanley Deser and Joel Franklin,
  ``Schwarzschild and Birkhoff \emph{a la} Weyl'',\\
  Am.\ J.\ Phys.\  {\bf 73} (2005) 261
  [\href{https://arxiv.org/abs/gr-qc/0408067}{arXiv:gr-qc/0408067 [gr-qc]}].
  
 \bibitem{Ravndal} 
   Nils Voje Johansen, Finn Ravndal, 
   ``On the discovery of Birkhoff's theorem'', 
   Gen.Rel.Grav. 38 (2006) 537-540  
   [\href{https://arxiv.org/abs/physics/0508163}
   {arXiv:physics/0508163 [physics.hist-ph]}].

   
\bibitem{Skakala}
J.~Skakala and M.~Visser,\\
``Birkhoff-like theorem for rotating stars in (2+1) dimensions'',\\{}
[\href{https://arxiv.org/abs/0903.2128}{arXiv: 0903.2128 [gr-qc]}].
   

\bibitem{painleve1}
Paul Painlev\'e, 
``La m\'ecanique classique et la th\'eorie de la relativit\'e\,", \\
C. R. Acad. Sci. (Paris) 173, 677--680(1921).

\bibitem{painleve2}
Paul Painlev\'e, 
 ``La gravitation dans la m\'ecanique de Newton et dans la m\'ecanique d'Einstein", 
 C.R Acad. Sci. (Paris) 173, 873-886(1921).
 
\bibitem{gullstand}
Gullstrand, Allvar (1922).
``Allgemeine L\"osung des statischen Eink\"orperproblems in der Einsteinschen Gravitationstheorie". 
 Arkiv f\"or Matematik, Astronomi och Fysik. 16 (8): 1--15.
 
 
 \bibitem{doran}
C.~Doran,
``A New form of the Kerr solution'',
Phys. Rev. D \textbf{61} (2000), 067503
doi:10.1103/PhysRevD.61.067503
[\href{https://arxiv.org/abs/gr-qc/9910099}{arXiv:gr-qc/9910099 [gr-qc]}].

\bibitem{river}
A.~J.~Hamilton and J.~P.~Lisle,
``The River model of black holes'',\\
Am. J. Phys. \textbf{76} (2008), 519-532
doi:10.1119/1.2830526\\{}
[\href{https://arxiv.org/abs/gr-qc/0411060}{arXiv:gr-qc/0411060 [gr-qc]}].

\bibitem{natario}
J.~Natario,
``Painlev\'e-Gullstrand Coordinates for the Kerr Solution'',\\
Gen. Rel. Grav. \textbf{41} (2009), 2579-2586
doi:10.1007/s10714-009-0781-2\\{}
[\href{https://arxiv.org/abs/0805.0206}{arXiv:0805.0206 [gr-qc]}].

\clearpage
\bibitem{martel}
K.~Martel and E.~Poisson,\\
``Regular coordinate systems for Schwarzschild and other spherical space-times'',\\
Am. J. Phys. \textbf{69} (2001), 476-480
doi:10.1119/1.1336836\\{}
[\href{https://arxiv.org/abs/gr-qc/0001069}{arXiv: gr-qc/0001069 [gr-qc]}].

\bibitem{faraoni}
V.~Faraoni and G.~Vachon,
``When Painlev\'e-Gullstrand coordinates fail'',\\
Eur. Phys. J. C \textbf{80} (2020) no.8, 771,
doi:10.1140/epjc/s10052-020-8345-4\\{}
[\href{https://arxiv.org/abs/2006.10827}{arXiv: 2006.10827 [gr-qc]}].




\bibitem{unexpected}
M.~Visser,\\
``Acoustic propagation in fluids: An unexpected example of Lorentzian geometry'',\\{}
[\href{https://arxiv.org/abs/gr-qc/9311028}{arXiv: gr-qc/9311028[gr-qc]}].

\bibitem{visser:1997}
M.~Visser,
``Acoustic black holes: Horizons, ergospheres, and Hawking radiation'',
Class. Quant. Grav. \textbf{15} (1998), 1767-1791
doi:10.1088/0264-9381/15/6/024
[\href{https://arxiv.org/abs/gr-qc/9712010}{arXiv:gr-qc/9712010 [gr-qc]}].

\bibitem{visser:1998}
M.~Visser,
``Acoustic black holes'',
[\href{https://arxiv.org/abs/gr-qc/9901047}{arXiv:gr-qc/9901047 [gr-qc]}].

\bibitem{volovik:1999}
G.~Volovik,
``Simulation of Painlev\'e-Gullstrand black hole in thin He-3-A film'',
JETP Lett. \textbf{69} (1999), 705-713
doi:10.1134/1.568079
[\href{https://arxiv.org/abs/gr-qc/9901077}{arXiv:gr-qc/9901077 [gr-qc]}].

\enlargethispage{50pt}
\bibitem{stone:2001}
S.~E.~Perez-Bergliaffa, K.~Hibberd, M.~Stone and M.~Visser,
``Wave equation for sound in fluids with vorticity'',
Physica D \textbf{191} (2004), 121-136
doi:10.1016/j.physd.2003.11.007
[\href{https://arxiv.org/abs/cond-mat/0106255}{arXiv:cond-mat/0106255 [cond-mat]}].

\bibitem{visser:2001}
M.~Visser, C.~Barcel\'o and S.~Liberati,
``Analog models of and for gravity'',\\
Gen. Rel. Grav. \textbf{34} (2002), 1719-1734
doi:10.1023/A:1020180409214\\{}
[\href{https://arxiv.org/abs/gr-qc/0111111}{arXiv:gr-qc/0111111 [gr-qc]}].

\bibitem{fischer:2002}
U.~R.~Fischer and M.~Visser,
``On the space-time curvature experienced by quasiparticle excitations in the 
Painlev\'e--Gullstrand effective geometry'',\\
Annals Phys. \textbf{304} (2003), 22-39
doi:10.1016/S0003-4916(03)00011-3
[\href{https://arxiv.org/abs/cond-mat/0205139}{arXiv:cond-mat/0205139 [cond-mat]}].

\bibitem{novello:2002}
M.~Novello, M.~Visser and G.~Volovik,
{\sl Artificial black holes},\\
(World Scientific, Singapore, 2002)

\bibitem{probing}
C.~Barcel\'o, S.~Liberati and M.~Visser,
``Probing semiclassical analog gravity in Bose-Einstein condensates with widely tunable interactions'',\\
Phys. Rev. A \textbf{68} (2003), 053613
doi:10.1103/PhysRevA.68.053613\\{}
[\href{https://arxiv.org/abs/cond-mat/0307491}{arXiv:cond-mat/0307491 [cond-mat]}].

\bibitem{equatorial-vortex}
M.~Visser and S.~Weinfurtner,
``Vortex geometry for the equatorial slice of the Kerr black hole,''
Class. Quant. Grav. \textbf{22} (2005), 2493-2510
doi:10.1088/0264-9381/22/12/011
[\href{https://arxiv.org/abs/gr-qc/0409014}{arXiv:gr-qc/0409014 [gr-qc]}].

\bibitem{LRR}
C.~Barcel\'o, S.~Liberati and M.~Visser,
``Analogue gravity'',
Living Rev. Rel. \textbf{8} (2005), 12
doi:10.12942/lrr-2005-12
[\href{https://arxiv.org/abs/gr-qc/0505065}{arXiv:gr-qc/0505065 [gr-qc]}].

\bibitem{visser:2010}
M.~Visser and C.~Molina-Par\'is,
``Acoustic geometry for general relativistic barotropic irrotational fluid flow'',
New J. Phys. \textbf{12} (2010), 095014
doi:10.1088/1367-2630/12/9/095014
[\href{https://arxiv.org/abs/1001.1310}{arXiv:1001.1310 [gr-qc]}].


\clearpage
\bibitem{prado1}
P.~Mart\'in-Moruno and M.~Visser,
``Generalized Rainich conditions, generalized stress-energy conditions, and the Hawking-Ellis classification'',\\
Class. Quant. Grav. \textbf{34} (2017) no.22, 225014
doi:10.1088/1361-6382/aa9039
[\href{https://arxiv.org/abs/1707.04172}{arXiv:1707.04172 [gr-qc]}].

\bibitem{prado2}
P.~Mart\'in-Moruno and M.~Visser,
``Essential core of the Hawking--Ellis types'',\\
Class. Quant. Grav. \textbf{35} (2018) no.12, 125003
doi:10.1088/1361-6382/aac147
[\href{https://arxiv.org/abs/1802.00865}{arXiv:1802.00865 [gr-qc]}].


\bibitem{exact}
H.~Stephani, D.~Kramer, M.~MacCallum, C.~Hoenselaers, and E.~Herlt,\\
\emph{Exact solutions of Einstein's equations} (second edition),\\
(Cambridge University Press, Cambridge, 2003)



\bibitem{phenomenology}
R. Carballo-Rubio, F. Di Filippo, S. Liberati and M. Visser,\\
``Phenomenological aspects of black holes beyond general relativity'',\\
Phys. Rev. D \textbf{98} (2018) 124009.
[\href{https://arxiv.org/abs/1809.08238}{arXiv:1809.08238 [gr-qc]}].

\bibitem{pandora}
R.~Carballo-Rubio, F.~Di Filippo, S.~Liberati and M.~Visser,\\
``Opening the Pandora's box at the core of black holes'',\\
Class. Quant. Grav. \textbf{37} (2020) no.14, 145005
doi:10.1088/1361-6382/ab8141\\{}
[\href{https://arxiv.org/abs/1908.03261}{arXiv:1908.03261 [gr-qc]}].

\bibitem{complete}
R.~Carballo-Rubio, F.~Di Filippo, S.~Liberati and M.~Visser,\\
``Geodesically complete black holes'',
Phys. Rev. D \textbf{101} (2020), 084047
doi:10.1103/PhysRevD.101.084047
[\href{https://arxiv.org/abs/1911.11200}{arXiv:1911.11200 [gr-qc]}].

\bibitem{viability}
R. Carballo-Rubio, F. Di Filippo, S. Liberati, C. Pacilio and M. Visser,\\
``On the viability of regular black holes'',
JHEP \textbf{2018} (2018).\\{}
[\href{https://arxiv.org/abs/1805.02675}{arXiv:1805.02675} [gr-qc]].

\enlargethispage{60pt}
\bibitem{LISA}
E.~Barausse, E.~Berti, T.~Hertog, S.~A.~Hughes, P.~Jetzer, P.~Pani, T.~P.~Sotiriou, N.~Tamanini, H.~Witek, K.~Yagi, N.~Yunes, \emph{et al.},\\
``Prospects for Fundamental Physics with LISA'',\\
Gen. Rel. Grav. \textbf{52} (2020) no.8, 81
doi:10.1007/s10714-020-02691-1.
[\href{https://arxiv.org/abs/2001.09793}{arXiv:2001.09793 [gr-qc]}].

\end{thebibliography}
\end{document}